\documentstyle[aps,epsf,axodraw]{revtex}
\begin{document}
\draft
\twocolumn[\hsize\textwidth\columnwidth\hsize\csname@twocolumnfalse%
\endcsname
\title{Polymer Shape Anisotropy and the Depletion Interaction}
\author{Mario Triantafillou and Randall D. Kamien}
\address{Department of Physics and Astronomy, University of Pennsylvania,
Philadelphia, PA \ 19104, USA}
\date{19 October 1998; revised 7 December 1998}
\maketitle

\begin{abstract}
We calculate the second and third virial coefficients of the
effective sphere-sphere interaction due to polymer-induced depletion forces. By
utilizing the anisotropy of a typical polymer conformation, we can
consider polymers that are roughly the same size as the
spheres.  We argue that recent experiments are laboratory evidence 
for polymer shape anisotropy.
\end{abstract}
]

Over sixty years ago, Kuhn studied the conformations of polymer chains
\cite{Ku} and recognized that typical conformations of ideal chains
are {\sl not} spherically symmetric.
The intuitive idea of a
symmetric shape is a result of the isotropic 
end-to-end vector distribution of a random walk\cite{Do}:  
spherical symmetry results from
implicit rotational averaging of the polymer. 
In fact, a typical 
polymer conformation is anisotropic, with an aspect ratio
of roughly $3.4:1$ \cite{So}.

Nonetheless, there is little laboratory evidence for this asphericity.  
In solution, polymers rotate
randomly and there are no significant polymer-polymer correlations. 
Thus small angle light scattering measurements of polymer
solutions 
can only determine the average principal axis of the polymer shape.
If
there were a way to induce strong orientational correlations, 
light-scattering measurements could
discern an asymmetry.  We will show that the induced attraction
between regions of depleted polymer concentration (inclusions)
is a probe of this shape anisotropy.

The shape tensor characterizes the spatial distribution of monomers:
\begin{equation}
\label{shapetensor}
M_{\alpha\beta} = \int_0^N dn\, \left( R_\alpha(n) - \bar{R}_\alpha \right)
\left( R_\beta(n) - \bar{R}_\beta \right),
\end{equation}
where $R_\alpha(n)$ is the position of monomer $n$, $\alpha$ and $\beta$
label the Cartesian co\"ordinates
and
$\bar{R}_\alpha\equiv{1\over N}\int_0^N dn\, R_\alpha(n)$ 
is the polymer center of mass.  The polymer radius of gyration
is simply $R_G^2={\rm Tr}(M)$. 
Moreover, the eigenvalues of $M_{\alpha\beta}$,
$\lambda_1^2\le\lambda_2^2\le\lambda_3^2$, are the
average squared radii of gyration along the principal axes of inertia.   
Simulations \cite{So,Kr} have determined that
the the most likely shape has
\begin{equation}
\label{ratio}
\lambda_3^2:\lambda_2^2:\lambda_1^2\approx 11.8:2.7:1.0
\end{equation}
Indeed, the shape asymmetry exists and is rather large
\cite{Foot}.
Exploiting this shape anisotropy
as a calculational tool is the main theoretical component
of this letter.

Though hard spheres only interact by direct contact,
entropic effects of other particles present
can induce long-range interactions.
These sorts of forces are responsible for liquid-crystalline order
in lyotropic systems \cite{fraden} and
surface crystallization in hard-sphere fluids \cite{dinsmore}.
It is instructive to consider the
virial expansion
for a gas of identical balls.  Around each sphere of radius $r$
there is a sphere of radius $2r$ from
which the centers of the other spheres are excluded.
When two spheres are close
their excluded regions overlap leaving more free volume for the
remaining spheres and hence a larger entropy.  
Thus the propensity for  ``close'' configurations
can be interpreted as an entropic ``depletion'' force.

Recently \cite{verma}, monodisperse polymers (specifically, $\lambda$-phage
DNA) have been used to induce depletion forces between polystyrene spheres.
To model this system at low concentrations, one might replace the polymers with
spheres of radius $R_G$, the polymer radius of gyration.
Asakura and Oosawa derived a simple formula for the potential
between two large spheres of diameter $\sigma$ in a gas of smaller spheres of
diameter $D$.  
The Asakura--Oosawa (AO) potential is
\cite{As}:
\begin{equation}
\label{AO}
{U(R)\over k_{\scriptscriptstyle B}T} = {\Pi v\lambda^3 \over(\lambda - 1)^3}
\left[ 1 - \frac{3}{2} \left( \frac{R}{\sigma \lambda} \right) + \frac{1}{2}
\left( \frac{R}{\sigma \lambda} \right)^3 \right],
\end{equation}
where $R$ is the distance between the centers of the large spheres, $v$
is the volume of a small sphere, $\Pi$ is the osmotic pressure of the
small-sphere gas and $\lambda \equiv 1 + D/\sigma$.
The above approximation simply sets $D=2R_G$.
This is obviously a crude approximation to the true system.  A complete
analysis
at length-scales longer than
the polymer persistence length ($50 {\rm nm}$) would count
the number of self-avoiding random walks which avoid the two polystyrene
spheres. 

In general, the effective potential is of the form $U_{\rm eff} = \Pi V(R)$
where $\Pi$ is the osmotic pressure and $V$ is an $R$ dependent
recovered volume.  At low concentrations the osmotic pressure is not
adjustable: $\Pi=k_{\scriptscriptstyle B}Tc$.  The physics all lies
in $V(R)$.  The AO model gives a one-parameter family of functions,
depending on the
effective hard-sphere diameter $D$.    In principle
one could derive a virial expansion for this potential with each
term involving the evaluation of a set of cluster integrals, each
of which involves integrations over polymer degrees of freedom.  
We will derive a different
one-parameter family based on the known shape distribution of polymers and
argue that the data in \cite{verma} is the first laboratory evidence
of the conformational anisotropy.

We start by calculating the classical configurational integral $Q_N(R)$, the sum of
Boltzmann weights over all conformations of $N$ polymers with
two spherical inclusions separated by $R$.
The sphere-sphere effective potential $U_{\rm eff}(R)$ is:
\begin{equation}
\label{distribution}
P(R) = \frac{Q_N(R)}{\int d{\bf R} Q_N(R)} \equiv
{\exp\left\{-\beta U_{\rm eff}(R)\right\}\over
\int d{\bf R} \exp\left\{-\beta U_{\rm eff}(R)\right\}}.
\end{equation}
Although $Q_N(R)$ has terms which are independent of
$R$, the resulting effective potential $U_{\rm eff}(R)$ does not:
they cancel between numerator and denominator in
(\ref{distribution}).

To calculate $Q_N(R)$, we sum over all the
conformations and placements of $N$ polymers with the excluded-volume
Boltzmann weights: $1$ if the
polymers and the spheres do not overlap and $0$ otherwise.
We split the integration over each polymer into three parts.  The
first is an integration over the center of mass, the second an integration
over all rigid rotations of the conformation, while the third is the
remaining integration over ``internal'' degrees of freedom.  This final
integral is over each unique conformation -- two conformations are equivalent
if one is merely a rigid rotation or translations of the other.  We will
only integrate over one representative from each equivalency class.
Denoting the space of all such ``internal'' polymer conformations
by $\Upsilon$, we have:
\begin{equation}
\label{configurational}
Q_N =
 \frac{1}{N!}\prod_{i=1}^N \int_{\Upsilon}  dr_{i,\rm int}\int d{\bf r}_i
d\Omega_i
\, e^{-\beta U},
\end{equation}
where $dr_{i,\rm int}$ is the measure on the space of conformations for polymer $i$,
${\bf r}_i$ is its center of mass and $\Omega_i$ is its rigid rotation.  
We further divide the integration over $\Upsilon$
by characterizing each polymer conformation by its
principal axes.  Defining $g(\lambda_1,\lambda_2,\lambda_3)$ is the number
of conformations with axes $\lambda_i$, we have
\begin{eqnarray}
\label{configmoments}
&Q_N& = \frac{1}{N!} \prod_{i=1}^N \nonumber\\
&&\quad \int d\lambda_{i1} d\lambda_{i2}
d\lambda_{i3}
\, g(\lambda_{i1},\lambda_{i2},\lambda_{i3})
\int d{\bf r}_i d\Omega_i \, e^{-\beta U}.
\end{eqnarray}
To pass from (\ref{configurational}) to (\ref{configmoments}) we 
assumed that the internal degrees of freedom did not affect the
interaction potential $U$.  This is, of course, not precisely correct.  Our
approximation replaces each polymer by a solid ellipsoid, and
then considers the potential due to only to this shape.  While this
certainly removes many degrees of freedom, it includes more
degrees of freedom than replacing the polymers by spheres.
The proof of
the pudding shall be in the eating -- we will see that this approximation
is valid by comparison with data \cite{verma}.
Thus our approximation replaces the monomer--sphere and monomer--monomer 
potential with a sum
of pairwise, ellipsoid-ellipsoid or sphere-ellipsoid terms.   
Each term is infinite for any overlap and zero
otherwise.

We now reduce the complexity
of the integration in (\ref{configmoments}) by using the shape distribution
of polymers.  Since the distribution of
polymer shapes is peaked around the prolate spheroid, we {\sl only}
consider those polymer shapes.  This approximation does not
account for the entire space of principal axes, though we believe that
it does characterize the polymer conformations better than
a sphere.  More importantly, our approximation allows us to consider
polymers which are roughly the same size as the included spheres.
This has a great advantage when comparing to the experiments of Verma,
{\sl et al} \cite{verma}.  By comparison, most work in this field has 
treated the inclusions approximately while correctly modelling the medium
as a gas of random walkers.
In \cite{Ei} field theoretic methods were used to obtain the depletion potential.  There,
in order to 
calculate reliably, the authors considered the interaction
between inclusions much smaller than the polymers, {\sl i.e.} $\sigma\ll
R_G$.  In the opposite extreme,
one might consider polymers which are much smaller
than the spheres.  In this case, it is appropriate to study the induced Casimir
force between two walls and, to then approximate the interaction
between spheres via the Derjaguin approximation \cite{degennes}.
By approximating the polymers
as prolate spheroids, we can treat the inclusions exactly -- a great advantage
when the two extreme limits are not applicable.
\vbox{\begin{figure}
\epsfxsize=2.5truein
\centerline{\epsfbox{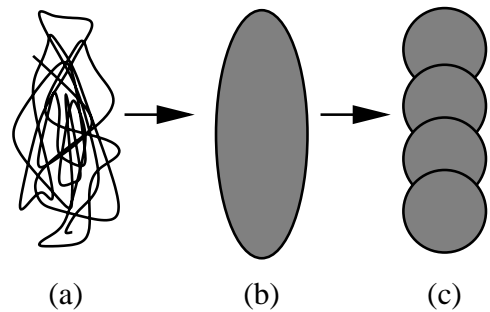}}
\vskip 8pt
\caption{Sequence of approximations. The actual polymer performs
a self-avoiding random walk (a) which has a typical prolate spheroidal shape
(b).
For computational simplicity, we neglect the anisotropy between the
two shorter principal axes and build up the resulting ellipsoid out of
overlapping spheres (c).}
\end{figure}}

We have thus reduced the configurational integral $Q_N$ over all
polymer modes to an integral over the allowed locations and
orientations of a prolate spheroid.  In order to reduce the phase space
somewhat and reduce our computation time,
we will assume that $\lambda_1=\lambda_2$.  
In the case of interest, we 
consider a prolate spheroid with aspect ratio $\sqrt{11.8}:1\approx
3.4:1.0$, so that
\begin{eqnarray}
\label{g=delta}
g(&\lambda_1,&\lambda_2,\lambda_3) \propto \nonumber\\
&&\delta (\lambda_1-\lambda_2)
\delta\big(\lambda_3 - (3.4)\lambda_1\big)\delta\big(R_G^2 - [2\lambda_1^2
+\lambda_3^2]\big),
\end{eqnarray}
where $R_G$ is the radius of gyration of the polymer.
To facilitate the numerical evaluation of integrals, we construct
the ellipsoid out of overlapping spheres.  Figure 1 depicts our sequence of
approximations. 
We note that choosing $\lambda_3\gg \lambda_2=\lambda_1$ would reduce our analysis
to that in \cite{Ya}
which considered
the depletion force between two spheres induced by thin
rods.  

Writing $f_{ij}=e^{-\beta u_{ij}}-1$, where $u_{ij}$ is
the excluded-volume potential between particles $i$ and $j$, and labelling
the spheres $A$ and $B$, $Q_N(R)$ can be rewritten as
\begin{eqnarray}
\label{configapprox}
&Q_N &= \frac{1}{N!} e^{-\beta u_{AB}} \int_V
d{\bf r}_1 d\Omega_1 d{\bf r}_2 d\Omega_2 \ldots d{\bf r}_N d\Omega_N \times
\\
 & & (1 + f_{A1})(1 + f_{B1})(1 + f_{A2})(1 + f_{B2})(1 + f_{12})
\ldots\nonumber
\end{eqnarray}
The product in \ref{configapprox} is a sum of
terms which can be grouped by the number of polymer positions and
rotations that are freely integrated over.  The first term is proportional
to $(4\pi V)^N$, the configurational integral for free ellipsoids, 
where $V$ is the volume of space minus
the volume of the two included spheres.  Subsequent terms 
have fewer powers of $V$.  Since the polymers are identical, these corrections
include combinatoric factors involving $N$. 
We take the limit $N\rightarrow\infty$ and $V\rightarrow\infty$ keeping $c\equiv N/V$ 
constant to
find the virial expansion in $c$.

It is both convenient and instructive
to graphically represent these ``cluster'' integrals \cite{Pathria} in
terms of Mayer cluster graphs.
The first two terms in $U_{\rm eff}(R)$ are:
\begin{eqnarray}
\label{second virial coefficient}
\thinlines
\beta &&U_{\rm eff}(R) = {c\over 4\pi}\int d{\bf r}_1 d\Omega_1 \,
\def\Base{\multiput(2,1)(8,0){2}{\circle{2}}}
\def\Dot{\put(6,8){\circle*{2}}}
\begin{picture}(14,15)
\Base \Dot \Line(2.5,1.87)(5.5,7.13) \Line(9.5,1.87)(6.5,7.13)
\end{picture} + {1\over 2}\left({c\over 4\pi}\right)^2 \\
 & & \times\int d{\bf
r}_1 d{\bf r}_2 d\Omega_1 d\Omega_2 \left[
\def\Base{\multiput(2,1)(8,0){2}{\circle{2}}}
\def\DoubleDot{\multiput(2,8)(8,0){2}{\circle*{2}}}
\def\Fac{\Line(2,2)(2,7)}
\def\Fad{\Line(2.7,1.7)(9.3,7.3)}
\def\Fbc{\Line(9.3,1.7)(2.7,7.3)}
\def\Fbd{\Line(10,2)(10,7)}
\def\Fcd{\Line(3,8)(9,8)}
\begin{picture}(125,15)
2 \Base \DoubleDot \Fac \Fbd \Fcd
\hspace{0.14in} +\,2 \Base \DoubleDot \Fac \Fbc \Fcd
\hspace{0.14in} +\,2 \Base \DoubleDot \Fac \Fbc \Fbd \Fcd
\hspace{0.14in} +\,2 \Base \DoubleDot \Fac \Fad \Fbc \Fcd
\hspace{0.14in} +\,\Base \DoubleDot \Fac \Fad \Fbc \Fbd \Fcd
\end{picture}
\right]\nonumber
\end{eqnarray}
where the open dots represent the spherical inclusions and the closed dots
represent the ellipsoids.
The integrals in (\ref{second virial coefficient})
are difficult to compute analytically and
thus were evaluated numerically via a Monte Carlo algorithm:
$10^3$ different angles
and $10^4$ different points were chosen in a volume which included both
spheres and which did not exclude any possible orientation or location of the
ellipsoids.  We calculated these integrals
by this random
sampling weighted by the appropriate phase-space volume factor.

To compare with experiment \cite{verma}, we took the
sphere diameter to
be $D=1.2\mu \rm m$ and $R_G = 0.5 \mu \rm m$.  Knowing $R_G$ enables us to
find the length of the ellipsoid, $L = 3.4\times2R_G /\sqrt{13.8} \approx
0.92 \mu \rm m$.  
We have chosen the hard-ellipsoid radii to be the mean
square radii of gyration, which is possibly na\"\i ve:
the hard-ellipsoid size should only be proportional
to $R_G$.  Since, in a random walk
the density decays as
$1/r$ there is no natural length scale which cuts off
the excluded-volume interaction.  Indeed, light scattering experiments
\cite{Na} have found that the effective
hard sphere radius
is roughly half the radius of gyration.  The relation between
$R_G$ and the hard-ellipsoid size must be determined through
the depletion-force experiments we are modeling.  

In Figure 2 we plot our
results as a function of concentration.  Until
one considers concentrations near the polymer overlap concentration
$c^*=1/(4\pi R_G^3/3)\approx 2/\mu{\rm m}^3$ the third virial coefficient, responsible
for a repulsive ``anti-correlation hole'', is a
small
perturbation to the leading term in (\ref{second virial coefficient}).  Thus
if we restrict our study to the dilute polymer regime, the leading term
in the virial expansion is sufficient. 
\vbox{\begin{figure}
\epsfxsize=3.0truein
\centerline{\epsfbox{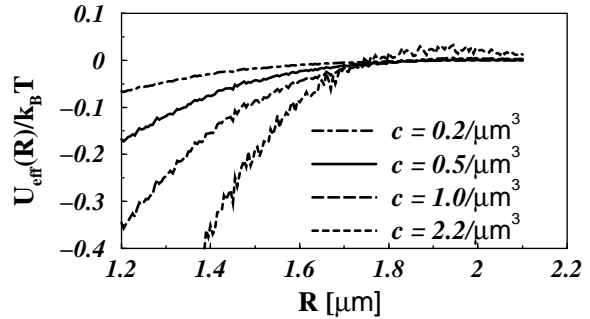}}
\vskip 8pt
\caption{Numerical result for the effective potential $U_{\rm eff}$ for
differing concentrations and $R_G = 0.5 \mu{\rm m}$.
We have kept terms up to second order in the
polymer concentration.  Note that at the highest concentration there
is a small repulsive bump at $R\approx 2\mu\rm m$.
However, for $c\le c*\approx 2/\mu{\rm m}^3$, the
third virial coefficient is a small contribution to $U_{\rm eff}$.}
\label{varryc}
\end{figure}}

We compare our model with the data \cite{verma} and the AO model
at $c=0.5/\mu{\rm m}^3$ for two reasons.  This concentration
is below the overlap concentration $c^*$ and at the same time
is large enough that the well depth is on the order of $k_{\scriptscriptstyle
B}T$ so that data can be reliably obtained.  We can
adjust the effective radius $R_G$ 
for both the AO model and our ellipsoid-based model.  
We find that a good fit results for the AO model
with $R_G=0.42 \mu\rm m$ and for our model with $R_G=0.8 \mu\rm m$, in
comparison with the light-scattering-obtained value $R_G=0.5 \mu\rm m$.
In Figure 3 we show the data along with these two one-parameter fits.  
We have checked other concentrations and have
found that at $c=1.0/\mu{\rm m}^3$ the theory and experiment also compare
favorably with the same effective radii.  
What should one conclude from this agreement?  It is possible
to interpret the data as arising from either shape simply by adjusting the
size of the shape.  However, one should start
from a microscopic picture based on the polymer
physics of the allowed chain conformations.  Only from this perspective
can one properly interpret the data.  

Since the radius of
gyration of
the $\lambda$-DNA can be calculated from its molecular weight and
persistence length to be $R_G=0.5
\mu\rm m$ we can consider two different zero parameter fits: our model and
the AO model.  The result is shown in Figure 4.  Note that the data lie
between our calculation and the AO model, suggesting that one could
smoothly
deform the AO sphere into an ellipsoid and find a best fit aspect ratio
to fit the data, giving a one-parameter fit.  Indeed, we have made a number of limited
runs on the fully anisotropic shape satisfying (\ref{ratio}).  This
results in a curve which is roughly $30\%$ deeper than that in Figure 4, which
is better than the simple sphero-cylinder result.
Finally, we have made similar comparisons
between theory and experiment 
at $c=0.1/\mu{\rm m}^3$, $c=0.2/\mu{\rm m}^3$ and $c=1.0/\mu{\rm m}^3$.  We find
that at $c=1.0/\mu{\rm m}^3$ the comparison is similar to that shown in Figure 4.
At the two lower concentrations, where the data is difficult to collect, 
neither our model nor the AO model
make very good predictions.

\vbox{\begin{figure}
\epsfxsize=3.0truein
\centerline{\epsfbox{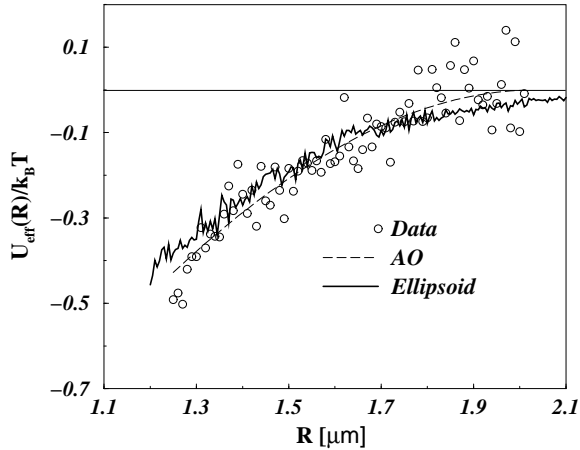}}
\vskip 8pt
\caption{ Our model and the AO model are fit to the experimental data for
$c = 0.5/ \mu \rm m^3$.  We varied only the effective $R_G$ as a parameter.
The fit gives $R_G = 0.42$ and $0.8$ for the AO model and our model
respectively.}
\label{Bag+AO+data}
\end{figure}}

\vbox{\begin{figure}
\epsfxsize=3.0truein
\centerline{\epsfbox{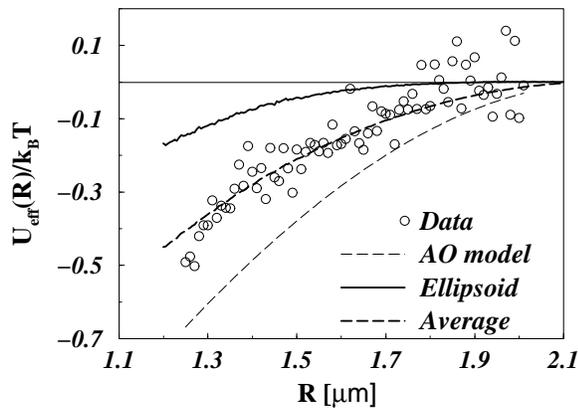}}
\vskip 8pt
\caption{Comparison of the AO model, the model presented here and the data at
$c=0.5/\mu{\rm m}^3$.  For both our model and the AO model we have taken
the theoretical value of $R_G \approx 0.5 \mu\rm m$.  There are no free
parameters
in the models.  We also plot the average of the AO model and the ellipsoid
calculation.}
\label{series of concentrations}
\end{figure}}  

Returning to the original question, are polymer shapes anisotropic?  Though
the full distribution $g(\lambda_1,\lambda_2,\lambda_3)$ is peaked, it has a
finite width. We can incorporate this width by choosing an appropriately
weighted admixture of shapes. 
Note that we could easily get a very good fit just by mixing spheres
and ellipsoids -- the mixed second virial coefficient is just a linear
combination of the two ``pure'' coefficients.  It is easy to see from
Figure 4 that
a 50-50 admixture of spheres and ellipsoids would give a remarkably good
fit to the data, without any adjustable parameters.  While we have no basis
for this weighting of shapes, 
we nonetheless take this as direct evidence for the anisotropy of
polymer conformations.

In closing, we note that while counting polymer conformations by
treating them as rigid ellipsoids is appropriate for static
properties, it is not clear at all that the dynamics reflects this.  In
particular one might ask whether a polymer ellipsoid rotates to a new orientation
slower or faster than it deforms into that orientation.  
Finally, our analysis could also be used 
to study the depletion interaction by actual ellipsoidal objects,
such as bacteria \cite{discher}.

It is a pleasure to acknowledge conversations with J.~Crocker, D.~Discher,
A.~Levine, T.~Lubensky, D.~Pine, R.~Verma and A.~Yodh.  We additionally
thank the authors of \cite{verma} for providing us with their data.
This research was supported in part by an award from Research Corporation, the
Donors of The Petroleum Research Fund,
administered by the American Chemical Society
and the NSF-MRSEC Program through Grant DMR96-32598.


\begin{thebibliography}{99}


\bibitem{Ku} W. Kuhn, Kolloid Z. {\bf 68}, 2 (1934).

\bibitem{Do} M. Doi and S.F. Edwards,  {\sl The Theory of Polymer Dynamics},
p. 29, (Oxford University Press, Oxford, 1986).

\bibitem{So} K. \v{S}olc, J. Chem. Phys. {\bf 55}, 335 (1971).


\bibitem{Kr} D.E. Kranbuehl and P.H. Verdier, J. Chem. Phys. {\bf 67}, 361.
(1977).

\bibitem{Foot}We note that although polymer self-avoidance
leads to a swelling of the polymer chain, it does not
significantly alter the shape anisotropy.  See J.A.  Aronovitz and D.R. Nelson, J. Phys. (Paris) {\bf 47}, 1445
(1986).

\bibitem{fraden} M.~Adams, Z. Dogic, S.L.~Keller and S. Fraden, Nature {\bf
393},
349 (1998).

\bibitem{dinsmore} A.D. Dinsmore, A.G. Yodh, D.J. Pine, Nature {\bf 383}, 239
(1996).

\bibitem{As} S. Asakura and F. Oosawa, J. Chem. Phys. {\bf 22}, 1255 (1954)

\bibitem{verma} R. Verma, J. Crocker, T.C. Lubensky and A.G. Yodh, Phys.
Rev. Lett., {\bf 81}, 404 (1998).

\bibitem{Ei} A. Hanke, E. Eisenriegler and S. Dietrich, preprint 
[cond-mat/9808225].

\bibitem{degennes} J.F. Joanny, L. Leibler and P.-G. de Gennes, J. Polym. Sci.,
Polym. Phys. Ed. {\bf
17}, 1073 (1979); see also Y. Mao,
M.E. Cates, and H.N.W. Lekkerkerker, J. Chem. Phys.
{\bf 106}, 3721 (1997).

\bibitem{Ya} K. Yaman, C. Jeppesen, and C.M. Marques, Europhys. Lett. {\bf
42}, 221 (1998).

\bibitem{Pathria} See, for instance, R.K. Pathria, {\sl Statistical Mechanics},
(Butterworth-Heinemann, New York, 1996).

\bibitem{Na}  X. Ye, T. Narayanan, P. Tong, and J.S. Huang, Phys. Rev. Lett.
{\bf 76}, 4640 (1996)

\bibitem{discher} We thank D.~Discher for discussions on this point.

\end{thebibliography}
\end{document}